# Multiscale modeling of femtosecond laser irradiation on copper film with electron thermal conductivity from *ab initio* calculation

Pengfei Ji and Yuwen Zhang[1]
Department of Mechanical and Aerospace Engineering
University of Missouri
Columbia, MO 65211, USA

**Abstract**

By combining *ab initio* quantum mechanics calculation and Drude model, electron temperature and lattice temperature dependent electron thermal conductivity is calculated and implemented into a multiscale model of laser material interaction, which couples the classical molecular dynamics and two-temperature model. The results indicated that the electron thermal conductivity obtained from *ab initio* calculation leads to faster thermal diffusion than that using the electron thermal conductivity from empirical determination, which further induces deeper melting region, larger number of density waves travelling inside the copper film and more various speeds of atomic clusters ablated from the irradiated film surface.

**Keywords:** Electron thermal conductivity, ab initio calculation, multiscale modeling, Drude model

## Nomenclature

| | |
|---|---|
| $A$ | material constants describing the electron-electron scattering rate, $s^{-1}K^{-2}$ |
| $B$ | material constants describing the electron-phonon scattering rate, $s^{-1}K^{-1}$ |
| $C_e$ | electron heat capacity, $J/(m^3 K)$ |
| $E$ | energy, $J$ |
| $f$ | Fermi-Dirac distribution function |
| $g$ | electron density of states |
| $G_{e-ph}$ | electron-phonon coupling factor, $W/(m^3 K)$ |
| $J$ | Laser fluence, $J/cm^2$ |
| $k$ | thermal conductivity, $W/(mK)$ |
| $k_B$ | Boltzmann constant, $1.38 \times 10^{-23} J/K$ |
| $L$ | penetrating depth, $m$ |
| $m$ | mass, $kg$ |
| $q$ | heat flux, $W/m^2$ |
| $\mathbf{r}_i$ | position of an nucleus |
| $R$ | reflectivity |
| $t$ | time, $s$ |
| $T$ | temperature, $K$ |
| $v$ | velocity, $m/s$ |
| $V_c$ | Volume of unit cell, $m^3$ |

**Greek Letters**

| | |
|---|---|
| $\varepsilon$ | electron energy level, $J$ |

---
[1] Corresponding author. Email: zhangyu@missouri.edu

| $\mu$ | chemical potential, $J$ |
| --- | --- |
| $\lambda\langle\omega^2\rangle$ | second moment of the electron-phonon spectral function, $meV^2$ |
| $\rho$ | density, $kg/m^3$ |
| $\tau_e$ | total electron scattering time |
| $\tau_{xx}$ | thermal stress, $GPa$ |

**Subscripts and Superscripts**

| $e$ | electron |
| --- | --- |
| $F$ | Fermi |
| $l$ | lattice |
| $op$ | optical |
| $p$ | pulse |

Acronyms and abbreviations widely used in text and list of references

| FDM | finite difference method |
| --- | --- |
| MD | molecular dynamics |
| QM | quantum mechanics |
| TTM | two-temperature model |

## 1. Introduction

In the past decades, fruitful progresses have been made in the field of laser interaction with metallic materials [1–4]. Due to significant differences between the masses of electron and nucleus, the electron-electron scattering time ($\sim 10fs$) is much shorter than the electron-phonon scattering time ($\sim 1ps$) [5]. When femtosecond laser irradiates on metal, the energy of laser pulse is firstly absorbed by electron subsystem, leading to the electrons to be heated to tens of thousands of degrees Kelvin. Whereas, the lattice temperature $T_l$ is still not fully heated during the time of femtosecond laser irradiation. In the subsequent tens of picoseconds, the deposited laser energy transfers from electron subsystem to the lattice subsystem. For a metal in the inertial confinement fusion context, solid to plasma phase transition induced by femtosecond laser pulse, is defined as warm dense matter. Besides laser material interaction, other applications, such as particle beam-target interaction, micromachining surface treatment, generation of plasma sources of X-rays also involve the investigation of warm dense matter [2,6,7]. A thorough understanding of the electron thermal conductivity $k_e$ of the warm dense matter is still under investigation.

Two-temperature method (TTM) is introduced to describe the energies of electron subsystem and lattice subsystem at the sates of their partial thermodynamic equilibriums, as well as the coupling of energy transfer between the two subsystems at the overall state of nonequilibrium [3]. The femtosecond laser interactions with gold film [8] and nanoparticle [9] were performed under the framework of TTM. The heat transfer process of femtosecond laser processing of metal was comprehensively studied by Chowdhury and Xu by establishing a parabolic two temperature model [10]. Different kinds of laser pulse duration were carried out to simulate the laser interaction with titanium [11]. It was found that the heat conduction effect in titanium was enhanced with the pulse duration increasing. During the heat transfer, a key parameter impacting the spatial degree of laser heating of the electron subsystem is the $k_e$. The heat conduction in an anisotropic thin film was studied by employing the coordinate transformation method [12], which showed that the anisotropic heat conduction was more significant under the condition of multiple consecutive pulses irradiation. In this letter, $k_{e,QM}$ is modeled and calculated by combining the *ab initio* quantum mechanics (QM) calculation with Drude model. A quantum mechanics, molecular dynamics and two-temperature model (QM-MD-TTM) integrated framework is constructed to investigate the impacts of $T_e$ and $T_l$ dependent $k_{e,QM}$ on femtosecond laser interaction with copper.

## 2. Modeling and Simulation

To illustrate the Drude model in a one–dimensional ($x$-direction) system [13], the energy of an electron at position $x$ is $E_x$. Before the collision of two electrons, their positions are at $x - \tau_e v_x$ and $x + \tau_e v_x$, respectively; here $\tau_e$ is the total scattering time of electron and $v_x$ is the velocity of electron in $x$- direction. The heat flux transport to $x$ is

$$q_x = \frac{1}{2} n_e [v_x E_{x-\tau_e v_x} + (-v_x) E_{x+\tau_e v_x}], \qquad (1)$$

where $n_e$ is the number of electrons per unit volume. By employing the chain rule to include electron temperature $T_e$, Equation (1) is rewritten as

$$q_x = -n_e v_x^2 \tau_e \frac{dE}{dT_e} \frac{dT_e}{dx}. \qquad (2)$$

Recalling Fourier's law $q_x = -k_e (dT_e/dx)$, the electron thermal conductivity $k_e$ becomes

$$k_e = n_e v_x^2 \tau_e \frac{dE}{dT_e}. \qquad (3)$$

For a three-dimensional metallic electron subsystem, when $T_e$ is lower than the Fermi temperature $T_F$, Eq. (3) becomes

$$k_e = \frac{1}{3} v_F^2 \tau_e \frac{dE}{dT_e}, \qquad (4)$$

where $v_F$ is the Fermi velocity. The total scattering rate of electron $\tau_e^{-1}$ is computed from the summation of electron-electron scattering rate $\tau_{e-e}^{-1} = AT_e^2$ and electron-phonon scattering rate $\tau_{e-ph}^{-1} = BT_l$, i.e., $\tau_e^{-1} = AT_e^2 + BT_l$, where the parameters $A$ and $B$ are two material constants [14]. The internal energy $E$ of electron subsystem is $\int_{-\infty}^{\infty} g f \varepsilon / V_c d\varepsilon$ [15], where $f = \{\exp[(\varepsilon - \mu)/k_B T_e] + 1\}^{-1}$ is the Fermi-Dirac distribution function. $\varepsilon$ and $\mu$ are the energy level and chemical potential of electron, respectively. $\mu$ varies with $T_e$ under electron excitation. $k_B$ is the Boltzmann constant. $g$ is the electron density of states (EDOS), which is a function of $T_e$ and $\varepsilon$. $V_c$ is the volume of the copper unit cell. Considering both $g$ and $f$ are electron temperature $T_e$ dependent variables, $dE/dT_e$ in Eq. (4) becomes

$$\frac{dE}{dT_e} = \frac{1}{V_c} \int_{-\infty}^{\infty} (\frac{\partial f}{\partial T_e} g + f \frac{\partial g}{\partial T_e}) \varepsilon d\varepsilon, \qquad (5)$$

Combing Eqs. (4) and (5), the $T_e$ and $T_l$ dependent $k_e$ is derived as

$$k_e = \frac{1}{V_c} \frac{v_F^2}{3(AT_e^2 + BT_l)} \int_{-\infty}^{\infty} (\frac{\partial f}{\partial T_e} g + f \frac{\partial g}{\partial T_e}) \varepsilon d\varepsilon. \qquad (6)$$

Since $k_e$ in Eq. (6) is calculated with the combination of QM and Drude model, it will be represented by $k_{e,QM}$ in the present work. In the calculation of $k_e$ in Eq. (6), the $T_e$ dependent $f$ and $g$ were obtained from *ab initio* calculation. The plane wave density functional theory code ABINIT [16] was utilized. The electron excitation at $T_e$ was represented by the finite temperature density functional theory proposed by Mermin [17]. A number of 11 valence electrons ($3d^{10}4s^1$) per atom were taken in the projector-augmented wave (PAW) atomic data. The exchange and correlation function was described by the local density approximation (LDA) [18]. After convergence test, the Brillouin zone was sampled by the Monkhorts-Pack method of $18 \times 18 \times 18$ $k$-point. The converged cut off energy was chosen as $32Ha$.

Besides the QM modeling of $k_e$ in the present work, the electron heat capacity $C_e$ [19] and the electron-phonon coupling factor $G_{e-ph}$ were also determined from QM modeling [20]

$$\begin{cases} C_e = \frac{1}{V_c} \int_{-\infty}^{\infty} (\frac{\partial f}{\partial T_e} g + f \frac{\partial g}{\partial T_e}) \varepsilon d\varepsilon \\ G_{e-ph} = \frac{1}{V_c} \frac{\pi \hbar k_B \lambda \langle \omega^2 \rangle}{g_{\varepsilon_F}} \int_{-\infty}^{\infty} g^2 (-\frac{\partial f}{\partial \varepsilon}) d\varepsilon' \end{cases} \quad (7)$$

where $\lambda \langle \omega^2 \rangle$ is the second moment of the electron-phonon spectral function, $g_{\varepsilon_F}$ is the number of EDOS at Fermi energy $\varepsilon_F$, both of which are $T_e$ dependent parameters. The QM part including electron excitation-dependent $k_e$, $C_e$ and $G_{e-ph}$, is coupled with the energy equation of electron subsystem in TTM

$$C_e \frac{\partial T_e}{\partial t} = \nabla(k_e \nabla T_e) - G_{e-ph}(T_e - T_l) + S(x,t), \quad (8)$$

where $S(x,t)$ is the energy source of laser pulse irradiating along the $x$- direction

$$S(x,t) = \frac{0.94J(1-R)}{t_p L_{op}} exp(-\frac{x}{L_{op}}) exp[-2.77 \frac{(t-t_0)^2}{t_p^2}], \quad (9)$$

where $J$ is the laser fluence. $R$ and $L_{op}$ denote the reflectivity and the optical penetration depth, respectively. $t_0$ is the temporal center point of the laser beam. $t_p$ is the full width of laser pulse at half maximum of laser intensity, which is defined as the laser pulse duration.

The lattice subsystem is modeled by using the classical molecular dynamics (MD). In one MD time step $\Delta t_{MD} = 1fs$ of the lattice subsystem, the finite difference method (FDM) computation of electron subsystem is performed $n_t$ times with a FDM time step of $\Delta_{FDM} = 0.005fs$ [21]. The von Neumann stability criterion [22] has to be met, namely,

$$\Delta t_{FDM} = \frac{\Delta t_{MD}}{n_t} < 0.5 \Delta x_{FDM}^2 \frac{C_e}{k_e} = 1.5 \frac{\Delta x_{FDM}^2}{v_F^2 \tau_e}. \quad (10)$$

The electron subsystem is divided into $N$ with $N_V$ atoms in the volume of $V_N$. At each time step of MD simulation, the energy transferring from electron subsystem to lattice subsystem for a given FDM cell is modeled as

$$E_{Transfer} = \frac{\Delta t_{MD}}{n_t} \sum_{k=1}^{n_t} G_{e-ph} V_N (T_e^k - T_l), \quad (11)$$

where $T_e^k$ is the average electron temperature in each $\Delta t_{MD}$. Therefore, the MD equation of lattice subsystem is

$$m_i \frac{d^2 r_i}{dt^2} = -\nabla U + \frac{E_{Transfer}}{\Delta t_{MD}} \frac{m_i v_i^T}{\sum_{j=1}^{N_V} m_j (v_j^T)^2}, \quad (12)$$

where $m_i$, $r_i$ and $v_i^T$ are the mass, position and thermal velocity of atom $i$. The embedded atom method (EAM) potential $U$ of copper, which was fitted and optimized from QM calculation [23], is adopted to describe the interatomic interaction in Eq. (13).

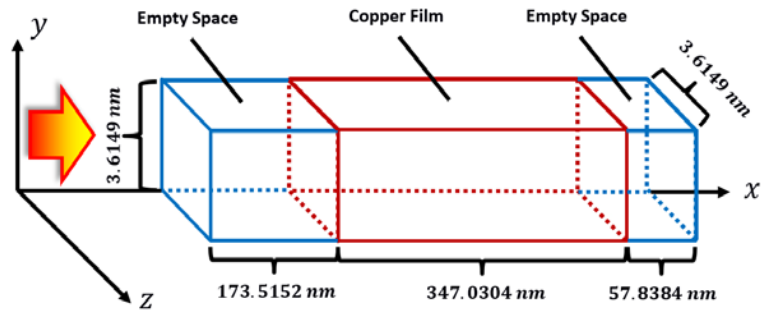

Fig. 1 Schematic show of the established system and direction of laser irradiation.

The QM-MD-TTM integrated framework is constructed by combing Eqs. (6), (7), (8) and (12). The numerical simulation is developed as an extension of the TTM part in the IMD [24] and the ABINIT code [16]. The established system is shown in Fig. 1, which is consisted with three components along the incidence of laser pulse ($x$- direction). The first and third components are two empty spaces, which are designed to allow expansion under the laser irradiation. The second component is the copper film. During the simulation process, periodic boundary conditions were applied in $y$- and $z$- directions of the system in Fig. 1. Free boundaries were set for the front and rear surfaces of the copper film. The entire simulation lasted for $100 ps$. The first $10\ ps$ was intended to make preparation before femtosecond laser irradiation, with $5\ ps$ of canonical ensemble (NVT) simulation to for equilibrating the system and another subsequent $5\ ps$ of micro-canonical ensemble (NVE) simulation for examining whether the equilibration had reached.

## 3. Results and Discussion
### 3.1. Comparisons of $k_e$ calculated from QM calculation and Empirical Determination

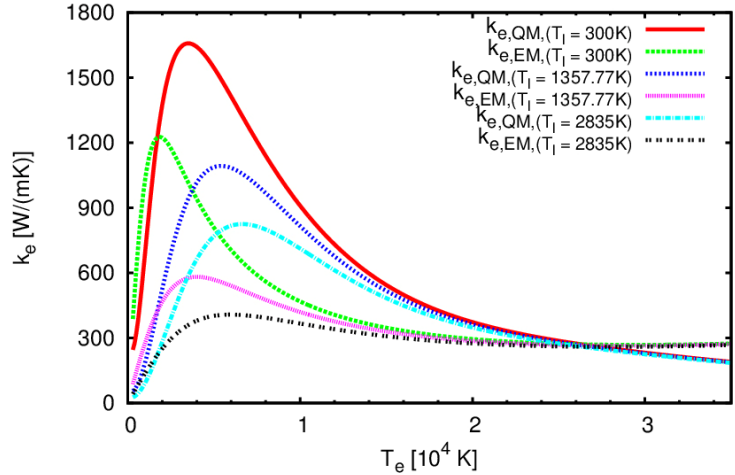

Fig. 2 Evolutions of electron thermal conductivity $k_{e,QM}$ from the QM calculation and $k_{e,EM}$ from empirical determination.

As derived in Eq. (7), $T_{e,QM}$ was calculated after obtaining the $T_e$ dependent $f$ and $g$ from *ab initio* QM calculation. For comparison, an empirically determined $k_{e,EM}$ [25], is plotted in Fig 2. The expression of $k_{e,EM}$ is shown below

$$k_{e,EM} = \chi \frac{(\vartheta_e^2 + 0.16)^{\frac{5}{4}}(\vartheta_e^2 + 0.44)\vartheta_e}{(\vartheta_e^2 + 0.092)^{\frac{1}{2}}(\vartheta_e^2 + \eta\vartheta_l)}, \qquad (13)$$

where $\vartheta_e = T_e/T_F$ and $\vartheta_l = T_l/T_F$. $T_F$ denotes the Fermi temperature of copper. $\chi$ and $\eta$ are constants [26]. The calculated $T_e$ dependent $k_{e,QM}$ and $k_{e,EM}$ at given $T_l$ are shown in Fig. 2. When $T_e < 2.7 \times 10^4 K$, $k_{e,QM}$ presents overall higher values than $k_{e,EM}$ at the given $T_l$. Both $k_{e,QM}$ and $k_{e,EM}$ shows steep increase since the initial points of laser heating. Peaks of $k_{e,QM}$ and $k_{e,EM}$ appear with the continuous increase of $T_e$. For the case of lower $T_l$, the peak starts earlier with higher values. When $T_l = 2,835K$, $k_{e,EM}$ shows the slightest increase and broadest peak. Even though $T_l$ are different, both $k_{e,QM}$ and $k_{e,EM}$ show converged trends when $T_e > 1 \times$ $10^4$, which indicates that $T_l$ is no longer the dominate factor that impacts the $k_e$. Analyzing Eq. (7), it can be seen that in the region of high $T_e$, $k_{e,QM}$ monotonically decreases with the increase of $T_e$, because the denominator $AT_e^2 + BT_l$ increases much greater than the numerator $\int_{-\infty}^{\infty}[(\partial f/\partial T_e)g + f(\partial g/\partial T_e)]\varepsilon d\varepsilon$. Therefore, the criterion of implementation of $k_{e,QM}$ is only applicable to be used in with $T_e < 3.5 \times 10^4 K$ (low $T_e$ region). Despite $k_{e,EM}$ was originally proposed to fit the wide range of $T_e$, it is empirically fitted and determined from experimental data.

### 3.2. Impact of $k_e$ on Femtosecond Laser Irradiation

The QM-MD-TTM integrated simulation is performed to investigate the impact of $k_e$ on the material response induced by femtosecond laser irradiation. Part of the QM-MD-TTM integrated simulation (by implementing $k_{e,EM}$) is carried out for comparison. The parameters used in the QM-MD-TTM integrated simulation is tabulated in Table 1. The temporal and spatial distribution of electron temperature $T_e$, lattice temperature $T_l$, density $\rho$ and thermal stress $\sigma_{xx}$ (along the $x$-direction) are calculated.

Table 1 Parameters in the QM-MD-TTM integrated simulation of laser interaction with copper.

| Parameter | Value |
| --- | --- |
| $v_F$ (m/s) | $1.57 \times 10^6$ [15] |
| $T_F$ (K) | $8.16 \times 10^4$ [15] |
| A (s$^{-1}$K$^{-2}$) | $1.75 \times 10^7$ [14] |
| B (s$^{-1}$K$^{-1}$) | $1.98 \times 10^{11}$ [14] |
| $L_{op}$ (nm) | 14.29 [26] |
| R | 0.4 [26] |
| J (J/cm$^2$) | 0.5 |
| N | 888 |
| $\Delta t_{MD}$ (fs) | 1 |
| $\Delta t_{FDM}$ (fs) | 0.005 |
| $n_t$ | 200 |
| $t_0$ (ps) | 15 |
| $t_p$ (fs) | 500 |
| $\chi$ (W/(mK)) | 377 [26] |
| $\eta$ | $\eta = 0.139$ [26] |

Figure 3 shows the spatial distributions of $T_e$ and $T_l$ at different time. The horizontal axis indicates the normalized position of the established system along $x$- direction. The regions from $0 < x < 0.3$ and $0.9 < x < 1.0$ are the two initially established empty spaces shown in Fig. 1. At the time scale of femtosecond, the energy transferring from electron subsystem to lattice subsystem is negligible, when it is compared with the thermal conduction inside the electron subsystem. It is why that the depths of heated regions (defined as $T_l > 500\ K$) are developed at $t_0 = 15\ ps$ in Fig. 3. For convenience of identifying the subfigures in Figs. 2-5, cases (a) and (b) represent the simulation results by taking $k_{e,QM}$ and $k_{e,EM}$, respectively. The region of laser heating (normalized position $0.3 < x < 0.7$) seen in Fig. 3(a) is deeper than the region (normalized position $0.3 < x < 0.5$) seen in Fig. 3(b), which can be explained by differences of thermal diffusivity $\alpha_{e,QM}$ and $\alpha_{e,EM}$. From the starting point (time $t < 15\ ps$) of a 500 $fs$ laser irradiation, $T_e$ and $T_l$ are at thermal equilibrium, which means $C_e$ are the same for the two cases. Due to the calculated $k_{e,QM}$ is greater than $k_{e,EM}$ (as shown in Fig. 2), and the negligible variations of density $\rho$ upon laser irradiation, it can be concluded that $\alpha_{e,QM}$ is greater than $\alpha_{e,EM}$. In other words, the thermal energy will be conducted faster for case (a) than that for case (b). When time lasts $\Delta t_{e,QM}$ and $\Delta t_{e,EM}$ are the same, the length of thermal energy travelling for the case using $k_{e,QM}$ will be longer. Therefore, Fig. 3(a) shows deeper region of laser heating than that of Fig. 3(b).

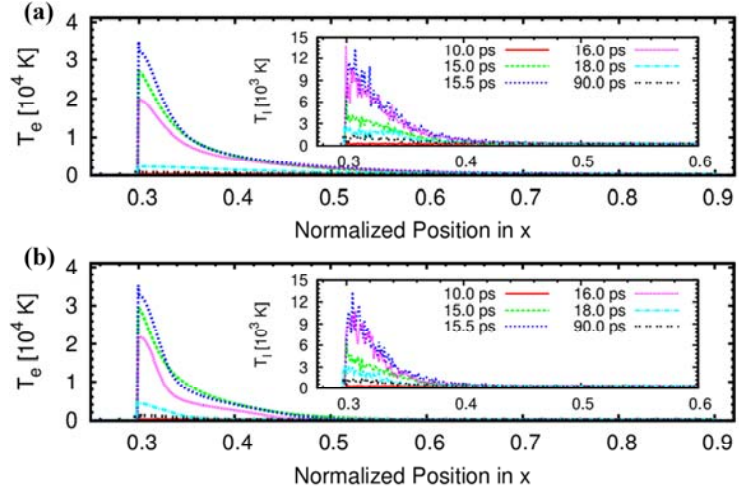

Fig. 3 Spatial distribution of electron temperature $T_e$ and lattice temperature $T_l$ by implementing (a) $k_{e,QM}$ and (b) $k_{e,EM}$.

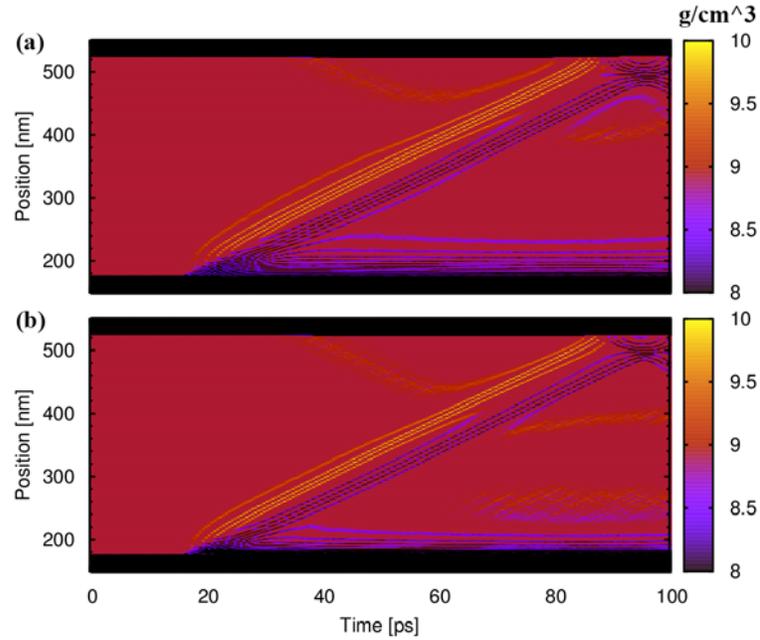

Fig. 4 Temporal and spatial distribution of density $\rho$ by implementing (a) $k_{e,QM}$ and (b) $k_{e,EM}$.

The temporal and spatial distribution of density $\rho$ provides a further insight into the internal structure change of copper film under laser irradiation. As illustrated in Fig. 4, the density of copper film before $15\ ps$ shows uniform $\rho_0$, which indicates that the copper atoms vibrate around their lattice points. The calculated $\rho_0$ at $T_l = 300\ K$ is $8.94\ g/cm^3$, which is slightly lower than the density of solid copper $8.985\ g/cm^3$ at $T_l = 293\ K$ [27]. Since laser irradiation ($\sim 15\ ps$), front side of the copper film presents smaller $\rho$ than $\rho_0$. As seen in the distributions of $\rho$ in the front side from $20\ ps$ to $40\ ps$, lower density appears in Fig. 4(b) than that in Fig. 4(a). Nevertheless, the waves (mixing with low and high density distributions) locating from $173.5152\ nm$ to $240\ nm$ in Fig. 4(a) shows deeper and denser than that in Fig. 4(b), which agrees with the depth of heated region for case (a) are deeper than that for case (b) discussed in the above paragraph. The lower density ($\rho < 8\ g/cm^3$) reflects that even though the depth of heated region is for case (b) is shallower, there are more thermal energy accumulates at the front surface of the copper film, which will results in greater local heating than that for case (b).

Right after laser irradiation, some of the waves stays at the front side of the copper film. Whereas, some of the waves travel along $x$- direction to the rear side the copper film. The number of dense waves ($\rho > 8.94\ g/cm^3$) appearing in Fig. 4(a) is greater than that appearing in Fig. 4(b). The number of tensile waves ($\rho < 8.94\ g/cm^3$) show the similar result with the compressible waves. The reasons can be attributed to higher degree of heated region for case (a) and case (b). Along the $x$-direction, the traveling speed of waves in Fig. 4(a) is slightly faster than those in Fig. 4(b).

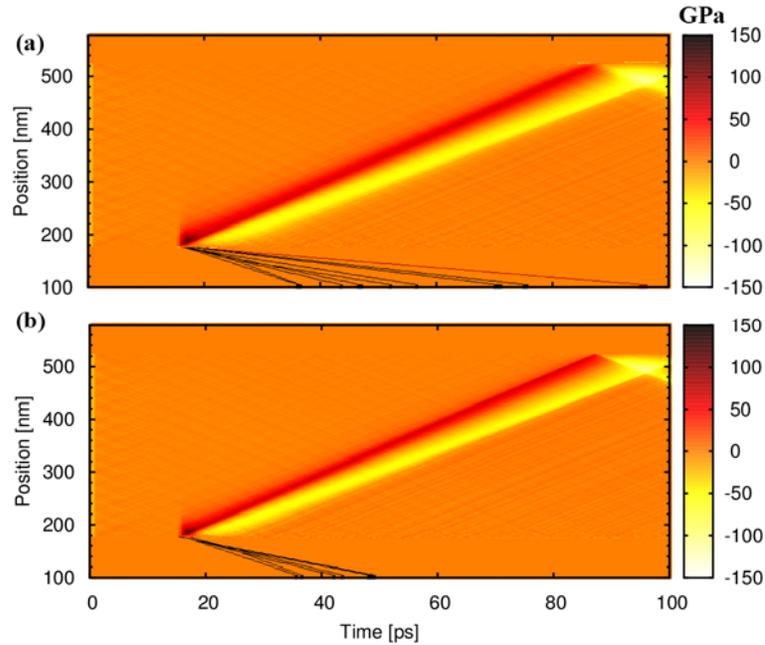

Fig. 5 Temporal and spatial distribution of thermal stress $\sigma_{xx}$ by implementing (a) $k_{e,QM}$ and (b) $k_{e,EM}$.

Besides the distribution of $\rho$, the temporal and spatial distribution of thermal stress $\sigma_{xx}$ is calculated by using the virial theorem [28] is drawn in Fig. 5. There are two sorts of stresses shown in Fig. 5. One is the stress $\sigma_{xx,cluster}$ of copper clusters, which ablate from the front surface of the copper film. The value of $\sigma_{xx,cluster}$ is as high as $150\ GPa$. The highest speed for these clusters incase (a) and case (b) are almost equal, which is ~$3.43\ km/s$. Recalling the spatial distribution of $T_l$ from $15.5\ ps$ to $16\ ps$ in the insets of Figs. 3(a) and 3(b), $T_l$ (near the front surface) in Fig. 3(a) is slightly greater than that in Fig. 3(b), which explains why there are clusters with more various speeds get ablated in cases (a) than those in case (b). The reason why these clusters are not detected and shown in Fig. 4 is that the densities of them are too small to be drawn in the range from $8\ g/cm^3$ to $10\ g/cm^3$. The other sort of the stress is the $\sigma_{xx,internal}$ traveling inside the copper film, whose travelling trajectory agrees with the wave of density distribution in Fig. 4. As computed in Fig. 5, $\sigma_{xx,internal}$ renders the traveling speed of ~$4.84\ km/s$, which is comparable to the longitudinal speed of sound for copper [29]. The traveling of stress inside metal films along the direction of incident laser pulse are also seen in [30,31].

## 4. Conclusion

In conclusion, $k_{e,QM}$ is modeled and determined by combing *ab initio* QM calculation with Drude method. The modeled $k_{e,QM}$ applies to condition of low $T_e$ region, which is a shortcoming of $k_{e,QM}$. The QM-MD-TTM integrated simulation results show that at given $T_e$ and $T_l$, greater $k_{e,QM}$ than $k_{e,EM}$ leads to faster thermal diffusion, deeper region of laser heating, more various speeds of atomic cluster ablations for case (a). The comparing the simulation results of implementing $k_{e,QM}$ and $k_{e,EM}$, provide a deeper insight into the role of electron thermal conductivity playing when a metal film is irradiated under femtosecond laser. Future work are suggested for the QM-MD-TTM integrated study the evaluation and mechanism of femtosecond laser interaction with metals, including melting, sublimation, and ablation phenomena.

## Acknowledgement

Support for this work by the U.S. National Science Foundation under grant number CBET-133611 is gratefully acknowledged.